\documentclass[a4paper]{article}
\usepackage{url, graphicx, color, varioref, amsfonts, longtable, float}
\usepackage{epstopdf,graphicx}
\usepackage{qcircuit}
\usepackage{algorithm}
\usepackage{algorithmic}

\newtheorem{theorem}{Theorem}[section]
\newtheorem{lemma}[theorem]{Lemma}

\newenvironment{proof}[1][Proof]{\begin{trivlist}
\item[\hskip \labelsep {\bfseries #1}]}{\end{trivlist}}

\newcommand{\qed}{\nobreak \ifvmode \relax \else
      \ifdim\lastskip<1.5em \hskip-\lastskip
      \hskip1.5em plus0em minus0.5em \fi \nobreak
      \vrule height0.75em width0.5em depth0.25em\fi}

\title{Reading a Single Qubit System Using Weak Measurement with Variable Strength}

\usepackage[T1]{fontenc}
\usepackage[utf8]{inputenc}
\usepackage{authblk}

\author[1,2]{Ahmed Younes\thanks{ayounes@alexu.edu.eg}}
\affil[1]{Department of Mathematics and Computer Science, Faculty of Science, Alexandria University, Egypt}
\affil[2]{School of Computer Science, University of Birmingham, Birmingham, B15 2TT, United Kingdom}
\renewcommand\Authands{ and }

\begin{document}
\maketitle

\begin{abstract}

Acquiring information about an unknown qubit in a superposition of two states is essential in any computation process. 
Quantum measurement, or sharp measurement, is usually used to read the information contents of that unknown qubit system. 
Quantum measurement is an irreversible operation that makes the superposition collapses to one of 
the two possible states in a probabilistic way. In this paper, a quantum algorithm will be proposed 
to read the information in an unknown qubit without applying sharp measurement on that qubit. 
The proposed algorithm will use a quantum feedback control 
scheme by applying sharp measurement iteratively on an auxiliary qubit weakly entangled with 
the unknown qubit. The information contents of the unknown qubit can be read by counting the outcomes from 
the sharp measurement on the auxiliary qubit. Iterative measurements on the auxiliary qubit will make the 
amplitudes of the superposition move in a random walk manner where a weak measurement is applied on the 
unknown qubit which can be reversed when the random walk takes opposite steps 
to decrease the disturbance introduced to the system. The proposed algorithm will define the strength 
of the weak measurement so that it can be controlled by adding an arbitrary number of dummy qubits $\mu$ to the system.
This will make the measurement process slowdown to an arbitrary scale so that the effect of the sharp measurement 
on the unknown qubit is reached after $O(\mu^2)$ measurements on the auxiliary qubit.

%
\end{abstract}

\section{Introduction}

Reading the information contents of an unknown qubit system is essential during any computation 
process, e.g. examining the contents and quantum error corrections. The reading process of a quantum system is usually 
done by measurements. Quantum Measurement, strong measurement, or sharp measurement is widely believed to be an irreversible \cite{R0}
operation that produce a probabilistic outcome by projecting the superposition of the possible states into a single state. 
Using strong measurement will destroy the original information contents of a qubit and might act as an error in this context. 

It was shown in \cite{R1,R2, R4} that a measurement process can be logically or physically reversible.
A measurement process is said to be logically reversible \cite{R1,R2} when the information about the pre-measurement state 
is preserved during the measurement \cite{R3} and can be recovered from the post-measurement state only if the post-measurement density operator and the outcome of the measurement 
can be used to fully calculate the pre-measurement density operator of the measured system, and so we can construct 
a logically reversible measurement for any sharp measurement that continuously approaches that sharp measurement 
with a decrease in the measurement error. 
A quantum measurement is said be physically reversible \cite{R2,R4} if the
pre-measurement state can be restored from the post-measurement
state in a probabilistic way using another reversing measurement so that 
the information about the system is preserved
during the measurement process and the original state
can be recovered using a physical process.

A physically reversible quantum measurement can be seen as a weak measurement where it was shown in \cite{R6,R7} that 
a quantum state post a partial-collapse measurement (weak measurement) can be recovered (uncollapsed) by adding 
a rotation and a second partial measurement with the same strength so that the extracted information from 
the partial-collapse measurement is erased, canceling the effect of both measurements. 
Physically reversible quantum measurement has been used in \cite{R5} 
on a spin-1/2 system using a spin-1/2 probe trying to completely specify an unknown quantum
state of a single system (see also erratum of Ref. \cite{R5}). 

Quantum feedback control was first studied in quantum optics \cite{R8,R9,R10}. Quantum feedback control was shown to have 
many applications, e.g. cooling an atom in an optical cavity \cite{R12}, 
measuring optical phase using adaptive measurements \cite{R13}, the stabilization of a single qubit, 
prepared in one of two nonorthogonal states against dephasing noise \cite{R11}, 
quantum error correction \cite{R17}, entanglement generation using measurement\cite{app16}, 
and quantum communication \cite{app18}.

It was shown in \cite{R10, R11, R14, R15, R16} that to obtain information about a quantum system, 
quantum feedback control using weak measurement can be used where the timescale of the measurement 
process can be extended where it takes the form of a random walk towards the final outcome such that 
the more the system is disturbed by the measurement, the more information is obtained about that system.

In this paper, a quantum algorithm will be 
proposed to acquire an unknown qubit system in order to obtain information about it without applying 
sharp measurement. The algorithm will read the content of that qubit using a quantum feedback control 
scheme where the sharp measurement on an auxiliary qubit will give the effect of weak measurement on the unknown 
qubit due to weak entanglement. The algorithm will make the amplitudes of 
the superposition move in a random walk manner to decrease the disturbance on the system where 
the opposite steps of the random walk will have a reversal effect on that system. 
The proposed algorithm will show that the strength of the weak measurement can be controlled 
by controlling the amount of disturbance introduced by adding an arbitrary number of dummy qubits to the system.
This can slowdown the measurement process to an arbitrary scale according to the amount of information needed such that the more we disturb the superposition, 
the more information we gain about it.

The paper is organized as follows: Section 2 defines the problem to be solved by the proposed algorithm. 
Section 3 defines the partial negation operator that will be used to create weak entanglement between 
the unknown qubit and an auxiliary qubit. Section 4 proposes the algorithm to read the information contents of an unknown qubit without applying 
sharp measurement on that qubit. Section 5 shows that weak measurement applied on the unknown qubit by applying iterative measurements 
on the auxiliary qubit has a reversal effect when the random walk moves in opposite directions. 
Section 6 shows that the algorithm will preserve the stability state so that the random walk converges 
to the correct destination even if the random walk moves up to some specific number of steps in the wrong direction. 
Section 7 defines the strength of the weak measurement and shows that this strength can be controlled 
based on the number of dummy qubits added to the system. 
Section 8 discusses the case of partial gain of information about the unknown qubit. 
The paper ends up with a conclusion in Section 9.

\section{Problem Statement}

Given a qubit $\left| \psi  \right\rangle$ with unknown $\phi$ as follows, 

\begin{equation}
\left| \psi  \right\rangle  = \cos \left( \phi  \right)\left| 0 \right\rangle  + \sin \left( \phi  \right)\left| 1 \right\rangle. 
\end{equation}

It is required to know how close the qubit to either $\left| 0 \right\rangle$ or $\left| 1 \right\rangle$ 
without too much disturbance to the superposition, i.e. no projective measurement is allowed on that qubit since projective measurement 
will make the qubit collapses to either $\left| 0 \right\rangle$ with probability  $\cos^2(\phi)$ or 
to $\left| 1 \right\rangle$ with probability  $\sin^2(\phi)$.

\section{Partial Negation Operator}

\begin{center}
\begin{figure*}[t]
\begin{center}

\setlength{\unitlength}{3947sp}%
\begingroup\makeatletter\ifx\SetFigFont\undefined
\def\x#1#2#3#4#5#6#7\relax{\def\x{#1#2#3#4#5#6}}%
\expandafter\x\fmtname xxxxxx\relax \def\y{splain}%
\ifx\x\y   
\gdef\SetFigFont#1#2#3{%
  \ifnum #1<17\tiny\else \ifnum #1<20\small\else
  \ifnum #1<24\normalsize\else \ifnum #1<29\large\else
  \ifnum #1<34\Large\else \ifnum #1<41\LARGE\else
     \huge\fi\fi\fi\fi\fi\fi
  \csname #3\endcsname}%
\else
\gdef\SetFigFont#1#2#3{\begingroup
  \count@#1\relax \ifnum 25<\count@\count@25\fi
  \def\x{\endgroup\@setsize\SetFigFont{#2pt}}%
  \expandafter\x
    \csname \romannumeral\the\count@ pt\expandafter\endcsname
    \csname @\romannumeral\the\count@ pt\endcsname
  \csname #3\endcsname}%
\fi
\fi\endgroup
\begin{picture}(2771,2237)(585,-1718)
\thinlines
\put(3002,-1494){\oval(84,92)}
\put(2678,-1514){\oval(236, 72)[tr]}
\put(2678,-1514){\oval(236, 72)[tl]}
\put(1646,-939){\circle*{90}}
\put(1278,-715){\circle*{90}}
\put(2206,-1270){\circle*{90}}

\put(1186,-1610){\framebox(176,215){}}
\put(1554,-1610){\framebox(176,215){}}
\put(1276,-686){\line( 0,-1){705}}
\put(943,-1505){\line( 1, 0){235}}
\put(1364,-1501){\line( 1, 0){188}}
\put(2112,-1614){\framebox(176,215){}}
\put(2288,-1501){\line( 1, 0){188}}
\put(2016,-935){\line( 1, 0){1328}}
\put(2016,-1277){\line( 1, 0){1328}}
\put(1824,-1273){\line(-1, 0){893}}
\put(1824,-935){\line(-1, 0){893}}
\put(1828,-716){\line(-1, 0){893}}
\put(2015,-1501){\line( 1, 0){ 94}}
\put(1737,-1499){\line( 1, 0){ 94}}
\put(2202,-1255){\line( 0,-1){141}}
\put(1644,-927){\line( 0,-1){470}}
\put(1098,-1706){\framebox(2118,1155){}}
\put(2482,-1397){\line( 0,-1){235}}
\put(2482,-1632){\line( 1, 0){376}}
\put(2858,-1632){\line( 0, 1){235}}
\put(2858,-1397){\line(-1, 0){376}}
\put(2567,-1572){\vector( 2, 1){282}}
\put(2866,-1496){\line( 1, 0){470}}
\put(2012,-716){\line( 1, 0){1316}}

\put(480,-1558){$\left| 0 \right\rangle$}
\put(480,-714){$\left| x_0 \right\rangle$}
\put(480,-989){$\left| x_1 \right\rangle$}
\put(480,-1319){$\left| x_{n-1} \right\rangle$}

\put(3385,-1558){$\left| ax \right\rangle$}

\put(956,-1180){$\vdots$}

\put(1215,-1584){$V$}
\put(1578,-1579){$V$}
\put(2139,-1593){$V$}

\put(1845,-1319){$\ldots$}
\put(1845,-989){$\ldots$}
\put(1845,-714){$\ldots$}
\put(1845,-1558){$\ldots$}

\put(1750,-475){$M_x$}

\end{picture}%

\end{center}
\caption{Quantum circuits for the $M_x$ operator followed by a partial measurement then 
reset the auxiliary qubit $\left| {ax} \right\rangle$ to state $\left| {0} \right\rangle$. }
\label{mmfig2}
\end{figure*}
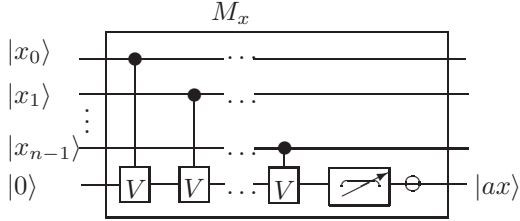
\end{center}

Let $X$ be the Pauli-X gate which is the quantum equivalent to 
the NOT gate. It can be seen as a rotation of the Bloch Sphere around the X-axis by $\pi$ radians as follows,

\begin{equation}
X  = \left[ {\begin{array}{*{20}c}
   0 & 1  \\
   1 & 0  \\
\end{array}} \right].
\end{equation}

The $c^{th}$ partial negation operator $V$ is the $c^{th}$ root of the $X$ gate and 
can be calculated using diagonalization as follows, 

\begin{equation}
 V=\sqrt[c]{X} = \frac{1}{2}\left[ {\begin{array}{*{20}c}
   {1 + t} & {1 - t}  \\
   {1 - t} & {1 + t}  \\
\end{array}} \right],
\end{equation}

\noindent
where $t={\sqrt[c]{{ - 1}}}$, and applying $V$ for $d$ times on a qubit is equivalent to the operator,
 
\begin{equation}
 V^d  = \frac{1}{2}\left[ {\begin{array}{*{20}c}
   {1 + t^d } & {1 - t^d }  \\
   {1 - t^d } & {1 + t^d }  \\
\end{array}} \right],
\end{equation}

\noindent
such that if $d=c$, then $V^d=X$. 

The $V$ gate will be used to define an operator $M_x$ as follows \cite{Younes}, 
$M_x$ is an operator on $n+1$ qubits register that applies $V$ conditionally 
for $n$ times on an auxiliary qubit initialized to state $\left|0 \right\rangle$ and will be denoted as $\left|ax \right\rangle$. 
The number of times the $V$ gate is applied on $\left|ax \right\rangle$ is based on the 1-density of a vector 
$\left| {x_0 x_1  \ldots x_{n-1} } \right\rangle$, where the 1-density of a state vector 
is the number of qubits in state ${\left| 1 \right\rangle }$, 
as follows (as shown in Fig. \ref{mmfig2}),

\begin{equation}
M_x = Cont\_V(x_0 ;ax )Cont\_V(x_1 ;ax ) \ldots Cont\_V(x_{n - 1} ;ax),
\end{equation}
\noindent
where the $Cont\_V(x_j ;ax )$ gate is a 2-qubit controlled gate with control qubit ${\left| x_j \right\rangle }$ and 
target qubit ${\left| ax \right\rangle }$. The $Cont\_V(x_j ;ax )$ gate applies $V$ conditionally on 
${\left| ax \right\rangle }$ if ${\left| x_j \right\rangle }= {\left| 1 \right\rangle }$, 
so, if $d$ is the 1-density of $\left| {x_0 x_1  \ldots x_{n-1} } \right\rangle$ then,

\begin{equation}
M_x\left( {\left| {x_0 x_1 ...x_{n - 1} } \right\rangle  \otimes \left| 0 \right\rangle } \right) = \left| {x_0 x_1 ...x_{n - 1} } \right\rangle  \otimes \left( {\frac{{1 + t^{d} }}{2}\left| 0 \right\rangle  + \frac{{1 - t^{d} }}{2}\left| 1 \right\rangle } \right),
\end{equation}
\noindent
and the probabilities of finding the auxiliary qubit $\left|ax \right\rangle$ in state 
${\left| 0 \right\rangle }$ or ${\left| 1 \right\rangle }$ when measured is respectively as follows,

\begin{equation}
\begin{array}{l}
 Pr{(\left| ax \right\rangle = \left| 0 \right\rangle)}  = \left| {\frac{{1 + t^d }}{2}} \right|^2  = \cos ^2 \left( {\frac{{d\pi }}{{2c}}} \right), \\ 
 Pr{(\left| ax \right\rangle = \left| 1 \right\rangle)}  = \left| {\frac{{1 - t^d }}{2}} \right|^2  = \sin ^2 \left( {\frac{{d\pi }}{{2c}}} \right). \\ 
 \end{array}
\end{equation}

\section{The Proposed Algorithm}

\subsection{Register Preparation}

Given an unknown qubit
$\left| \psi  \right\rangle  = \cos \left( \phi  \right)\left| 0 \right\rangle  + \sin \left( \phi  \right)\left| 1 \right\rangle$, 
append a quantum register of $\mu+1$ qubits to $\left| \psi  \right\rangle$ , 
where the $\mu$ qubits are all initialized to state $\left| 1 \right\rangle$ and a single auxiliary qubit
$\left| ax  \right\rangle$ initialized to state $\left| 0 \right\rangle$ as follows,

\begin{equation}
\begin{array}{l}
 \left| {\psi _{ext} } \right\rangle = \left| \psi  \right\rangle  \otimes \left| 1 \right\rangle ^{ \otimes \mu }  \otimes \left| 0 \right\rangle \\
  \,\,\,\,\,\,\,\,\,\,\,\,\,\,\,\,\,= \cos \left( \phi  \right)\left( {\left| 0 \right\rangle  \otimes \left| 1 \right\rangle ^{ \otimes \mu }  \otimes \left| 0 \right\rangle } \right) + \sin \left( \phi  \right)\left( {\left| 1 \right\rangle  \otimes \left| 1 \right\rangle ^{ \otimes \mu }  \otimes \left| 0 \right\rangle } \right) \\ 
 \,\,\,\,\,\,\,\,\,\,\,\,\,\,\,\,\, = \cos \left( \phi  \right)\left( {\left| {\psi _0 } \right\rangle  \otimes \left| 0 \right\rangle } \right) + \sin \left( \phi  \right)\left( {\left| {\psi _1 } \right\rangle  \otimes \left| 0 \right\rangle } \right). \\ 
 \end{array}
\end{equation}

The number of the $\mu$ qubits is a free parameter that will be used to adjust the accuracy of the proposed algorithm according 
to our purposes as will be shown later.

\subsection{The Algorithm}

When the operator $M_x$ is applied on $\left| {\psi _{ext} } \right\rangle$, it gives,

\begin{equation}
\begin{array}{l}
M_x \left| {\psi _{ext} } \right\rangle  = \cos \left( \phi  \right)\left( {\left| {\psi _0 } \right\rangle  \otimes \left( {\frac{{1 + t^{d_0 } }}{2}\left| 0 \right\rangle  + \frac{{1 - t^{d_0 } }}{2}\left| 1 \right\rangle } \right)} \right) \\
\,\,\,\,\,\,\,\,\,\,\,\,\,\,\,\,\,\,\,\,\,\,\,\,\,\,\,\,\, + \sin \left( \phi  \right)\left( {\left| {\psi _1 } \right\rangle  \otimes \left( {\frac{{1 + t^{d_1 } }}{2}\left| 0 \right\rangle  + \frac{{1 - t^{d_1 } }}{2}\left| 1 \right\rangle } \right)} \right),
\end{array}
\end{equation}
\noindent 
where $d_0$ is the 1-density of the state $\left| \psi_0 \right\rangle$ and 
$d_1$ is the 1-density of the state $\left| \psi_1 \right\rangle$, then $d_0=\mu$ and $d_1=\mu+1$, the probabilities of finding the auxiliary qubit $\left|ax \right\rangle$ in state 
${\left| 0 \right\rangle }$ or ${\left| 1 \right\rangle }$ when measured is respectively as follows,

\begin{equation}
\label{Pr_0_ax0}
 {\rm Pr_0}\left( {\left| {{\rm ax}} \right\rangle  = \left| {\rm 0} \right\rangle } \right) = \sin ^2 \left( \phi  \right)\cos ^2 \left( {\theta _1 } \right) + \cos ^2 \left( \phi  \right)\cos ^2 \left( {\theta _0 } \right),
\end{equation}

\begin{equation}
\label{Pr_0_ax1}
 {\rm Pr_0}\left( {\left| {{\rm ax}} \right\rangle  = \left| {\rm 1} \right\rangle } \right) = \sin ^2 \left( \phi  \right)\sin ^2 \left( {\theta _1 } \right) + \cos ^2 \left( \phi  \right)\sin ^2 \left( {\theta _0 } \right), \\ 
\end{equation}
\noindent
where $\theta_0=\frac{\pi d_0}{2c}$	and $\theta_1=\frac{\pi d_1}{2c}$.

\begin{algorithm}
\label{alg2}
\caption{Measurement Based Quantum Random Walk}
\begin{algorithmic}[1] 
\STATE Prepare $\left| {\psi _{ext} } \right\rangle$
\STATE Let $j_0=0$
\STATE Let $j_1=0$
\FOR {$counter = 1 \to r$}
	\STATE Apply the operator $M_x$ on $\left| {\psi _{ext} } \right\rangle$.
	\STATE Measure $\left| ax \right\rangle$
	\IF {$\left|ax \right\rangle=\left|1 \right\rangle$}
		\STATE $j_1=j_1+1$
	\ELSE
		\STATE  $j_0=j_0+1$
	\ENDIF
	\STATE Reset $\left|ax \right\rangle$ to state $\left|0 \right\rangle$
\ENDFOR
	\IF {$j_1>j_0$}
		\STATE The qubit $\left| {\psi} \right\rangle$ is closer to state $\left| 1 \right\rangle$ 
	\ELSE
		\STATE The qubit $\left| {\psi} \right\rangle$ is closer to state $\left| 0 \right\rangle$ 
	\ENDIF
\end{algorithmic}
\end{algorithm}

Applying the Algorithm on $\left| {\psi _{ext} } \right\rangle$ for $j \ge 1$ iterations with $j=j_0+j_1$, such that 
$j_0$ counts how many times we found $\left|ax \right\rangle=\left|0 \right\rangle$ and $j_1$ counts how many times 
we found $\left|ax \right\rangle=\left|1 \right\rangle$, then 
the amplitudes of the system will be updated after each iteration according to the following recurrence relations, 
let the system at iteration $j \ge 1$ is as follows,

\begin{equation}
\left| {\psi _{ext}^j } \right\rangle  = \alpha _j \left| {\psi _0 } \right\rangle  + \beta _j \left| {\psi _1 } \right\rangle, 
\end{equation}
\noindent
with $\alpha _0=\cos(\phi )$ and $\beta _0=\sin(\phi )$. The probability to find 
$\left|ax \right\rangle=\left|0 \right\rangle$ or $\left|ax \right\rangle=\left|1 \right\rangle$ is as follows,

\begin{equation}
 {\rm Pr_j} \left( {\left|ax \right\rangle=\left|0 \right\rangle} \right) = \alpha _j^2 \cos ^2 \left( {\theta _0 } \right) + \beta _j^2 \cos ^2 \left( {\theta _1 } \right),
\end{equation}

\begin{equation}
 {\rm Pr_j} \left( {\left|ax \right\rangle=\left|1 \right\rangle} \right) = \alpha _j^2 \sin ^2 \left( {\theta _0 } \right) + \beta _j^2 \sin ^2 \left( {\theta _1 } \right). 
\end{equation}

When measurement is applied on $\left| ax \right\rangle$, if we find $\left|ax \right\rangle=\left|0 \right\rangle$ then the amplitudes of the system will be updated as follows,

\begin{equation}
 \alpha _{j + 1}  = \frac{{\alpha _j \cos \left( {\theta _0 } \right)}}{{\sqrt {\Pr _j \left( {ax = 0} \right)} }},
\end{equation}
\begin{equation} 
 \beta _{j + 1}  = \frac{{\beta _j \cos \left( {\theta _1 } \right)}}{{\sqrt {\Pr _j \left( {ax = 0} \right)} }}, 
\end{equation}

\noindent
and if we find $\left|ax \right\rangle=\left|1 \right\rangle$ then the amplitudes of the system will be updated as follows,
\begin{equation}
 \alpha _{j + 1}  = \frac{{\alpha _j \sin \left( {\theta _0 } \right)}}{{\sqrt {\Pr _j \left( {ax = 1} \right)} }}, 
\end{equation}
\begin{equation}
 \beta _{j + 1}  = \frac{{\beta _j \sin \left( {\theta _1 } \right)}}{{\sqrt {\Pr _j \left( {ax = 1} \right)} }}. 
\end{equation}


The following equations are the closed forms of the above recurrence relations such that 
${\rm Pr_j}(\left| \psi_0 \right\rangle) = \alpha^2 _j$ and ${\rm Pr_j}(\left| \psi_1 \right\rangle) = \beta^2 _j$.
The probabilities of finding the auxiliary qubit $\left|ax \right\rangle$ in state 
${\left| 0 \right\rangle }$ or ${\left| 1 \right\rangle }$ when measured is respectively as follows,

\begin{equation}
\label{Pr_j_ax0}
 {\rm Pr_j}\left( {\left| {{\rm ax}} \right\rangle  = \left| {\rm 0} \right\rangle } \right) = \frac{{\sin ^2 \left( \phi  \right)\cos ^{2\left( {j_0  + 1} \right)} \left( {\theta _1 } \right)\sin ^{2j_1 } \left( {\theta _1 } \right) + \cos ^2 \left( \phi  \right)\cos ^{2\left( {j_0  + 1} \right)} \left( {\theta _0 } \right)\sin ^{2j_1 } \left( {\theta _0 } \right)}}{{\sin ^2 \left( \phi  \right)\cos ^{2j_0 } \left( {\theta _1 } \right)\sin ^{2j_1 } \left( {\theta _1 } \right) + \cos ^2 \left( \phi  \right)\cos ^{2j_0 } \left( {\theta _0 } \right)\sin ^{2j_1 } \left( {\theta _0 } \right)}} \\ 
\end{equation}

\begin{equation}
\label{Pr_j_ax1}
 {\rm Pr_j}\left( {\left| {{\rm ax}} \right\rangle  = \left| {\rm 1} \right\rangle } \right) = \frac{{\sin ^2 \left( \phi  \right)\cos ^{2j_0 } \left( {\theta _1 } \right)\sin ^{2\left( {j_1  + 1} \right)} \left( {\theta _1 } \right) + \cos ^2 \left( \phi  \right)\cos ^{2j_0 } \left( {\theta _0 } \right)\sin ^{2\left( {j_1  + 1} \right)} \left( {\theta _0 } \right)}}{{\sin ^2 \left( \phi  \right)\cos ^{2j_0 } \left( {\theta _1 } \right)\sin ^{2j_1 } \left( {\theta _1 } \right) + \cos ^2 \left( \phi  \right)\cos ^{2j_0 } \left( {\theta _0 } \right)\sin ^{2j_1 } \left( {\theta _0 } \right)}} \\ 
\end{equation}

\noindent
and the probabilities of states $\left| \psi_0  \right\rangle$ and $\left| \psi_1  \right\rangle$ will be changed 
according to the outcome of the measurement on $\left| {{ax}} \right\rangle$, i.e. $j_1$ will be incremented by 1 if 
$\left| {{ax}} \right\rangle= \left| {{1}} \right\rangle$, and  $j_0$ will be incremented by 1 if 
$\left| {{ax}} \right\rangle= \left| {{0}} \right\rangle$, so the probabilities of states 
$\left| \psi_0  \right\rangle$ and $\left| \psi_1  \right\rangle$ after $j\ge 1$ iterations will be as follows,

\begin{equation}
 {\rm Pr_j}\left( {\left| \psi_0  \right\rangle} \right) = \frac{{\cos ^2 \left( \phi  \right)\cos ^{2j_0 } \left( {\theta _0 } \right)\sin ^{2j_1 } \left( {\theta _0 } \right)}}{{\sin ^2 \left( \phi  \right)\cos ^{2j_0 } \left( {\theta _1 } \right)\sin ^{2j_1 } \left( {\theta _1 } \right) + \cos ^2 \left( \phi  \right)\cos ^{2j_0 } \left( {\theta _0 } \right)\sin ^{2j_1 } \left( {\theta _0 } \right)}} \\ 
 \label{prjpsi0}
 \end{equation}

\begin{equation}
 {\rm Pr_j}\left( {\left| \psi_1  \right\rangle} \right) = \frac{{\sin ^2 \left( \phi  \right)\cos ^{2j_0 } \left( {\theta _1 } \right)\sin ^{2j_1 } \left( {\theta _1 } \right)}}{{\sin ^2 \left( \phi  \right)\cos ^{2j_0 } \left( {\theta _1 } \right)\sin ^{2j_1 } \left( {\theta _1 } \right) + \cos ^2 \left( \phi  \right)\cos ^{2j_0 } \left( {\theta _0 } \right)\sin ^{2j_1 } \left( {\theta _0 } \right)}} \\ 
\label{prjpsi1}
 \end{equation}

The first aim of the algorithm is to make the measurement on $\left| ax  \right\rangle$ has a weak effect on 
the probabilities of $\left| \psi  \right\rangle$, i.e. weak measurement. This can be done by setting $c$ in $M_x$ such that 
$c>d_1$ so that finding $\left| ax  \right\rangle = \left| 0  \right\rangle$ will not make 
$\left| \psi_1  \right\rangle$ disappear from the superposition. One more benefit from using 
weak measurement is that weak measurement can be reversed as will proved later.

The second aim is to get $j_0>j_1$ with high probability if $\sin ^2 \left( \phi  \right) < \cos ^2 \left( \phi  \right)$, 
and vice versa, and since the value of $\phi$ is unknown, so we need to make 
${\rm Pr}\left( {\left| {{\rm ax}} \right\rangle = \left| {\rm 1} \right\rangle } \right)$ 
and ${\rm Pr}\left( {\left| {{\rm ax}} \right\rangle  = \left| {\rm 0} \right\rangle } \right)$ as close as possible to 0.5 so that 
the impact of $\phi$ appears on the probabilities of $\left| ax  \right\rangle$. Setting the probabilities of 
 $\left| ax  \right\rangle$ as close as possible to 0.5 will also make the measurement on $\left| ax  \right\rangle$ 
 has a small impact on the probabilities of $\left| \psi  \right\rangle$.
 
 To satisfy the above two aims, we need to set $\theta_0=\frac{\pi}{4}-\varepsilon$ and 
$\theta_1=\frac{\pi}{4}+\varepsilon$ for small $\varepsilon>0$. This can be done by setting the parameters as follows,

\begin{equation}
\begin{array}{l}
 d_0  = \mu, \\ 
 d_1  = \mu+1 , \\ 
 c = 2\mu+1,   \\ 
 \end{array}
\end{equation}
\noindent
so that,

\begin{equation}
\begin{array}{l}
\theta _0  = \frac{{\pi \mu }}{{2\left( {2\mu  + 1} \right)}}, \\
\theta _1  = \frac{{\pi \left( {\mu  + 1} \right)}}{{2\left( {2\mu  + 1} \right)}}.\\
\end{array}
\end{equation}

\section{Reversibility of Weak Measurement}
During the run of the proposed algorithm, repetitive measurement on $\left| ax  \right\rangle$ will slightly change 
the probabilities of ${\left| \psi_0  \right\rangle}$ and ${\left| \psi_1  \right\rangle}$. If after an arbitrary measurement, 
we find  $\left| ax  \right\rangle = \left| 0  \right\rangle$, then the probability of ${\left| \psi_0  \right\rangle}$ will 
increase, and if we find  $\left| ax  \right\rangle = \left| 1  \right\rangle$, then the probability of ${\left| \psi_1  \right\rangle}$ will 
increase. This section will show that after arbitrary number of measurements on $\left| ax  \right\rangle$, if the number of times 
we found $\left| ax  \right\rangle = \left| 0  \right\rangle$ equals to the number of times 
we found $\left| ax  \right\rangle = \left| 1  \right\rangle$, then the probabilities of ${\left| \psi_0  \right\rangle}$ 
and ${\left| \psi_1  \right\rangle}$ will be restored to the initial probabilities, i.e. 
finding $\left| ax  \right\rangle = \left| 0  \right\rangle$ after any measurement on $\left| ax  \right\rangle$ 
will reverse the effect of finding 
$\left| ax  \right\rangle = \left| 1  \right\rangle$ after any other measurement and vice versa. To prove this, 
we need the following lemma.

\begin{lemma}
\label{lemma1all}
Let $\theta _0  = \frac{{\pi \mu }}{{2\left( {2\mu  + 1} \right)}}$ and 
$\theta _1  = \frac{{\pi \left( {\mu  + 1} \right)}}{{2\left( {2\mu  + 1} \right)}}$ for any $\mu \ge 1$, 
then for any $m \ge 0$,

\begin{equation}
\frac{{\cos ^m \left( {\theta _1 } \right)\sin ^m \left( {\theta _1 } \right)}}{{\cos ^m \left( {\theta _0 } \right)\sin ^m \left( {\theta _0 } \right)}} = 1.
\label{lemma1}
\end{equation}

\begin{proof}

Since $\theta _0  = \frac{{\pi \mu }}{{2\left( {2\mu  + 1} \right)}}$ and 
$\theta _1  = \frac{{\pi \left( {\mu  + 1} \right)}}{{2\left( {2\mu  + 1} \right)}}$, then 
$\theta_0$ and $\theta_1$ can be re-written as,

\begin{equation}
\begin{array}{l}
 \theta _0  = \frac{\pi }{4} - \varepsilon, \\ 
 \theta _1  = \frac{\pi }{4} + \varepsilon,  \\ 
 \end{array}
\end{equation}

\noindent
with  $\varepsilon  = \frac{\pi }{{4\left( {2\mu  + 1} \right)}}$, then,

\begin{equation}
\begin{array}{l}
\label{cossin1}
 \cos \left( {\theta _1 } \right) = \cos \left( {\frac{\pi }{4} + \varepsilon } \right) \\ 
 \,\,\,\,\,\,\,\,\,\,\,\,\,\,\,\,\,\,\,\,\, = \frac{1}{{\sqrt 2 }}\left( {\cos \left( \varepsilon  \right) - \sin \left( \varepsilon  \right)} \right) \\ 
 \,\,\,\,\,\,\,\,\,\,\,\,\,\,\,\,\,\,\,\,\, = \sin \left( {\frac{\pi }{4} - \varepsilon } \right) \\ 
 \,\,\,\,\,\,\,\,\,\,\,\,\,\,\,\,\,\,\,\,\, = \sin \left( {\theta _0 } \right), \\ 
 \end{array}
\end{equation}

\noindent
and,

\begin{equation}
\begin{array}{l}
\label{cossin2}
 \sin \left( {\theta _1 } \right) = \sin \left( {\frac{\pi }{4} + \varepsilon } \right) \\ 
 \,\,\,\,\,\,\,\,\,\,\,\,\,\,\,\,\,\,\,\,\, = \frac{1}{{\sqrt 2 }}\left( {\cos \left( \varepsilon  \right) + \sin \left( \varepsilon  \right)} \right) \\ 
 \,\,\,\,\,\,\,\,\,\,\,\,\,\,\,\,\,\,\,\,\, = \cos \left( {\frac{\pi }{4} - \varepsilon } \right) \\ 
 \,\,\,\,\,\,\,\,\,\,\,\,\,\,\,\,\,\,\,\,\, = \cos \left( {\theta _0 } \right), \\ 
 \end{array}
\end{equation}

\noindent
and so Eqn.(\ref{lemma1}) holds.
\end{proof}
\end{lemma}

\begin{theorem}

Assume that the initial probabilities of ${\left| \psi_0  \right\rangle}$ and ${\left| \psi_1  \right\rangle}$ 
be  $\cos ^2 \left( \phi  \right)$ and $\sin ^2 \left( \phi  \right)$ respectively. Let $j_0$ and $j_1$ be the number 
of times we find ${\left| ax  \right\rangle}={\left| 0  \right\rangle}$ and  ${\left| ax  \right\rangle}={\left| 1  \right\rangle}$ 
respectively when measured.
If $j_0=j_1$ then the probabilities of ${\left| \psi_0  \right\rangle}$ and ${\left| \psi_1  \right\rangle}$ will be equal 
to the initial probabilities.

\begin{proof}

Assume that ${\left| ax  \right\rangle}$ is measured for $j$ times, where $j$ is an even number such that $j=j_0+j_1$ and $j\ge 0$. 
If $j_0=j_1$ then the proof holds directly using Lemma \ref{lemma1all} in Eqn.(\ref{prjpsi0}) and Eqn.(\ref{prjpsi1}).

\end{proof}

\end{theorem}

\section{Stability of the Proposed Algorithm}

Due to the symmetry of the problem, we can consider only the case when $\sin ^2 \left( \phi  \right) < \cos ^2 \left( \phi  \right)$, 
and the case of $\sin ^2 \left( \phi  \right) > \cos ^2 \left( \phi  \right)$ can be deduced by similarity.
It is clear from Eqns. (\ref{Pr_0_ax0}) and (\ref{Pr_0_ax1}) that before the first measurement on $\left| {{\rm ax}} \right\rangle$, we have 
${\rm Pr_0} \left({\left| ax  \right\rangle}={\left| 0  \right\rangle}\right) >  {\rm Pr_0} \left({\left| ax  \right\rangle}={\left| 1  \right\rangle}\right)$ 
if $\sin ^2 \left( \phi  \right) < \cos ^2 \left( \phi  \right)$, and from 
from Eqns. (\ref{Pr_j_ax0}) and (\ref{Pr_j_ax1}) we can see that the more we move in the correct direction, i.e. 
incrementing $j_0$ faster than $j_1$, the more we gain bias to ${\rm Pr_j} \left({\left| ax  \right\rangle}={\left| 0  \right\rangle}\right)$.

This section will show that even if the algorithm moves in the wrong direction, i.e. 
incrementing $j_1$ faster than $j_0$ when $\sin ^2 \left( \phi  \right) < \cos ^2 \left( \phi  \right)$, 
${\rm Pr_j} \left({\left| ax  \right\rangle}={\left| 0  \right\rangle}\right)$ will stay greater than 
${\rm Pr_j} \left({\left| ax  \right\rangle}={\left| 1  \right\rangle}\right)$ for a certain number of wrong measurements 
on $\left| ax  \right\rangle$, i.e. $\left| ax  \right\rangle = \left| 1  \right\rangle$, giving a high probability for the 
algorithm to recover from the effect of moving in the wrong direction.

 Given that $\cos \left( \theta_0 \right)= \sin \left( \theta_1 \right)$, and 
$\sin \left(  \theta_0 \right)= \cos \left(  \theta_1 \right)$ as shown in Eqns. (\ref{cossin1}) and (\ref{cossin2}), then the 
four master equations of the system shown in Eqns. (\ref{Pr_j_ax0}), (\ref{Pr_j_ax1}), (\ref{prjpsi0}) and (\ref{prjpsi1}) 
can be re-written as follows,

\begin{equation}
\label{Prjax0short}
 {\rm Pr_j} \left( {\left| {ax} \right\rangle  = \left| 0 \right\rangle } \right) = \frac{{\tan ^2 \left( \varphi  \right)\sin ^2 \left( {\theta _0 } \right) + \cos ^2 \left( {\theta _0 } \right)\tan ^{2 \Delta j} \left( {\theta _0 } \right)}}{{\tan ^2 \left( \varphi  \right) + \tan ^{2 \Delta j} \left( {\theta _0 } \right)}}, \\ 
\end{equation}
 \begin{equation}
\label{Prjax1short}
 {\rm Pr_j} \left( {\left| {ax} \right\rangle  = \left| 1 \right\rangle } \right) = \frac{{\tan ^2 \left( \varphi  \right)\cos ^2 \left( {\theta _0 } \right) + \sin ^2 \left( {\theta _0 } \right)\tan ^{2 \Delta j} \left( {\theta _0 } \right)}}{{\tan ^2 \left( \varphi  \right) + \tan ^{2 \Delta j} \left( {\theta _0 } \right)}}, \\ 
\end{equation}

\begin{equation}
\label{Prjpsi0short}
  {\rm Pr_j} \left( {\left| {\psi _0 } \right\rangle } \right) = \frac{{\tan ^{2 \Delta j} \left( {\theta _0 } \right)}}{{\tan ^2 \left( \varphi  \right) + \tan ^{2 \Delta j} \left( {\theta _0 } \right)}}, \\ 
\end{equation}

\begin{equation}
\label{Prjpsi1short}
  {\rm Pr_j} \left( {\left| {\psi _1 } \right\rangle } \right) = \frac{{\tan ^2 \left( \varphi  \right)}}{{\tan ^2 \left( \varphi  \right) + \tan ^{2 \Delta j} \left( {\theta _0 } \right)}}, \\ 
\end{equation}

\noindent
where $\Delta j = j_1  - j_0$. For the algorithm to be stable, then $\Delta j <0$ 
when $\sin ^2 \left( \phi  \right) < \cos ^2 \left( \phi  \right)$. We know that weak measurement is reversible, 
assume the random walk moves for $\Delta j >0 $ steps in the wrong 
direction. We need to know how far the random walk should go in the wrong direction while maintaining 
the stability condition $\Pr \left( {\left| {ax} \right\rangle  = \left| 0 \right\rangle } \right) > \frac{1}{2}$, so we get,

\begin{equation}
\sin ^2 \left( \phi  \right) < \cos ^2 \left( \phi  \right)\tan ^{2\Delta j} \left( {\theta _0 } \right),
\end{equation}

\noindent
such that, if $\Delta j = 0$, so we get the initial probabilities of the system, i.e.  
$\sin ^2 \left( \phi  \right) < \cos ^2 \left( \phi  \right)$, and we have 
$\Pr \left( {\left| {ax} \right\rangle  = \left| 0 \right\rangle } \right) > \frac{1}{2}$ as long as, 

\begin{equation}
\Delta j \ge \frac{{\log \left( {\tan \left( \phi  \right)} \right)}}{{\log \left( {\tan \left( {\theta _0 } \right)} \right)}} \ge 0.
\end{equation}
 
 This means that the algorithm will maintain the stability condition even if the random walk goes in the wrong 
direction for at most $\frac{{\log \left( {\tan \left( \phi  \right)} \right)}}{{\log \left( {\tan \left( {\theta _0 } \right)} \right)}}$ steps. 
This gives the algorithm a chance to restore the random walk to move in the correct direction

\section{The Strength of Weak Measurement}

The scale of a projective measurement is of length 1, so it has the maximum strength, 
after which the state of the unknown qubit will be 
projected to one of the eigen vectors of the system in a probabilistic way. 
The strength of the weak measurement can be understood as the distance that the random walk 
has to move from the initial state to the state that are  $\epsilon$-far from the projected state for small $\epsilon>0$. 
This section will show that the strength of the weak measurement can controlled by using 
an arbitrary number of dummy qubits $\mu$ in the system. It will be shown that 
the measurement process can be scaled to an arbitrary length based on the number of dummy qubits 
added to the system. 

Assuming again the case where $\sin ^2 \left( \phi  \right) < \cos ^2 \left( \phi  \right)$, 
then the scale of the measurement process is based upon the number of steps that the random walk should move 
starting from $\Pr _0 \left( {\left| {\psi _0 } \right\rangle } \right) = \cos ^2 \left( \varphi  \right)$ 
to reach after $j \ge 1$ steps to 
$\Pr _j \left( {\left| {\psi _0 } \right\rangle } \right) = 1 - \epsilon $ 
for small $\epsilon >0$, so

\begin{equation}
 {\rm Pr_j} \left( {\left| {\psi _0 } \right\rangle } \right) = \frac{{\tan ^{2\Delta j} \left( {\theta _0 } \right)}}{{\tan ^2 \left( \varphi  \right) + \tan ^{2\Delta j} \left( {\theta _0 } \right)}} \ge 1 - \epsilon, 
\end{equation}
\noindent

then,

\begin{equation}
\begin{array}{l}
 \Delta j \ge \frac{{\log \left( {\tan ^2 \left( \varphi  \right)\left( {\frac{{1 - \epsilon }}{\epsilon }} \right)} \right)}}{{\log \left( {\tan ^2 \left( {\theta _0 } \right)} \right)}} \\ 
\,\,\,\,\,\,\,\,\,\,  \ge \frac{{\log \left( {\tan ^2 \left( \varphi  \right)\left( {\frac{{1 - \epsilon }}{\epsilon }} \right)} \right)}}{{\log \left( {\sin ^2 \left( {\theta _0 } \right)} \right) - \log \left( {\cos ^2 \left( {\theta _0 } \right)} \right)}} \\ 
\,\,\,\,\,\,\,\,\,\,  \ge \frac{{\log \left( {\tan ^2 \left( \varphi  \right)\left( {\frac{{1 - \epsilon }}{\epsilon }} \right)} \right)}}{{\log \left( {\cos ^2 \left( {\theta _1 } \right)} \right) - \log \left( {\cos ^2 \left( {\theta _0 } \right)} \right)}} \\ 
 \,\,\,\,\,\,\,\,\,\, \ge \frac{{\log \left( {\tan ^2 \left( \varphi  \right)\left( {\frac{{1 - \epsilon }}{\epsilon }} \right)} \right)}}{{\log \left( {\cos ^2 \left( {\theta _1 } \right)} \right) - \log \left( {\cos ^2 \left( {\theta _0 } \right)} \right)}}, \\ 
 \end{array}
\end{equation}

\noindent
and since $\theta _0  = \frac{{\pi \mu }}{{4\mu  + 2}}$ and $\theta _1  = \frac{{\pi \left( {\mu  + 1} \right)}}{{4\mu  + 2}}$ then

\begin{equation}
\begin{array}{l}
 \Delta j \ge \frac{{\log \left( {\tan ^2 \left( \varphi  \right)\left( {\frac{{1 - \epsilon }}{\epsilon }} \right)} \right)}}{{\left( {\frac{{\pi \mu }}{{4\mu  + 2}}} \right)^2  - \left( {\frac{{\pi \left( {\mu  + 1} \right)}}{{4\mu  + 2}}} \right)^2 }} \\ 
 \,\,\,\,\,\,\,\,\,\, \ge \left( {\frac{2}{\pi }} \right)^2 \log \left( {\tan ^2 \left( \varphi  \right)\left( {\frac{{1 - \epsilon }}{\epsilon }} \right)} \right)\left( {2\mu  + 1} \right). \\ 
 \end{array}
\end{equation}

For suffiently large $\mu >0$,
$  {\rm Pr_j} \left( {\left| {ax} \right\rangle  = \left| 0 \right\rangle } \right) = \frac{1}{2} + \delta$ 
and $ {\rm Pr_j} \left( {\left| {ax} \right\rangle  = \left| 1 \right\rangle } \right) = \frac{1}{2} - \delta $ for small 
 $\delta>0$, then \cite{expdist}, 
  
\begin{equation}
\Delta j = \sqrt {\frac{2}{\pi }j} 
\end{equation}

\noindent
and since $\varphi$ is unknown, then assume $\varphi=\frac{\pi}{2}$ as an upper bound for 
the total number of steps $j$ and so the scale of the measurement process is,
\begin{equation}
\begin{array}{l}
 j_{proj} \ge \frac{\pi }{2}\left( {\Delta j} \right)^2  \\ 
\,\,\,\,\,\,\,\,\,\,  \ge \frac{2}{\pi }\left( {\log \left( {\tan ^2 \left( \frac{\pi}{2}  \right)\left( {\frac{{1 - \epsilon }}{\epsilon }} \right)} \right)\left( {2\mu  + 1} \right)} \right)^2  \\ 
\,\,\,\,\,\,\,\,\,\,  \ge O\left( {\mu ^2 } \right). \\ 
 \end{array}
\end{equation}

This means that if the algorithm is iterated for $j_{proj}$ iterations, then 
${\rm Pr_{proj} \left( j_0>j_1 \right)}= {\sin ^2 \left( \varphi  \right)}$ similar to the case of the 
projective measurement.

\section{Partial Gain of Information}

Assume the case when we are given a certain number of dummy qubits $\mu$ and we do not want to iterate 
the algorithm for $j_{proj}$ times, but we want to stop early at iteration $J<j_{proj}$ for not fully disturbing 
the superposition, then we need to find ${\rm Pr_J \left( j_0>j_1 \right)}$ after $J$ iterations.

When $\varphi<\frac{\pi}{4}$, the algorithm is assumed to be successful if $j_0>j_1 $ and vice verse. 
Without losing of generality, assume $J$ is even, then the algorithm is assumed successful 
when we read $\left| ax \right\rangle = \left| 0 \right\rangle$ for at least $\frac{J}{2}+1$ times, 
i.e.  $j_0>\frac{J}{2}$, then

\begin{equation}
\Pr \left( {j_0  > \frac{J}{2}} \right) = \sum\limits_{k = \frac{J}{2} + 1}^J {\left( {\begin{array}{*{20}c}
   J  \\
   k  \\
\end{array}} \right)} \left( {{\rm Pr _J} \left( {\left| {ax} \right\rangle  = \left| 0 \right\rangle } \right)} \right)^k \left( {\rm Pr _J \left( {\left| {ax} \right\rangle  = \left| 1 \right\rangle } \right)} \right)^{J - k},
\end{equation}
\noindent
where $
\left( {\begin{array}{*{20}c}
   J  \\
   k  \\
\end{array}} \right) = \frac{{J!}}{{k!(J - k)!}}
$, and we know that ${\rm Pr_J \left( j_0>j_1 \right)}=\frac{1}{2}$ as a trivial case when $\varphi=\frac{\pi}{4}$, i.e. when 
$\left| \psi  \right\rangle  = \frac{1}{{\sqrt 2 }}\left( {\left| 0 \right\rangle  + \left| 1 \right\rangle } \right)$,  
then the probability of success of the algorithm after $J$ iterations 
with $\Delta J = - \sqrt {\frac{2}{\pi }J}$ is as follows,

\begin{equation}
{\rm Pr _J} \left( {j_0  > j_1 } \right) = \sin ^2 \left( \varphi  \right) + \cos \left( {2\varphi } \right)\Pr \left( {j_0  > \frac{J}{2}} \right).
\label{psucc}
\end{equation}

\begin{center}
\begin{figure}[htbp]
\begin{center}
   \includegraphics[width=250pt]{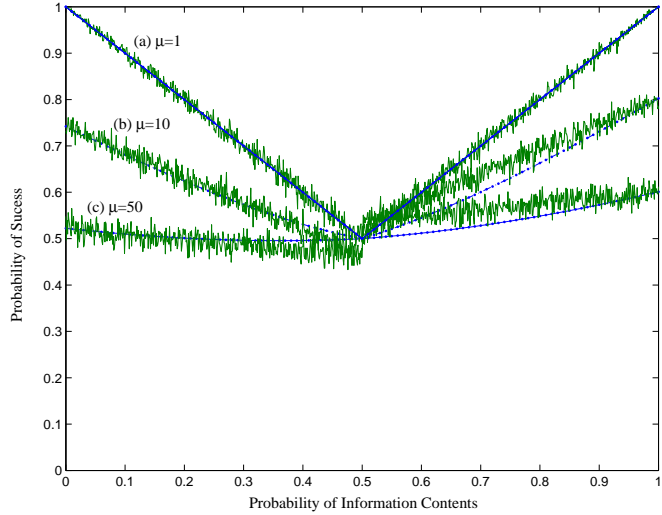}
   \caption{The probability of success for the measurement based quantum random walks with different values of $\mu$, 
   where the solid lines refers to the simulation results and the dotted lines is the probability of success 
shown in Eqn.(\ref{psucc}).}
   \label{readout}
\end{center}
\end{figure}
\end{center}

\begin{center}
\begin{figure}[htbp]
\begin{center}
   \includegraphics[width=250pt]{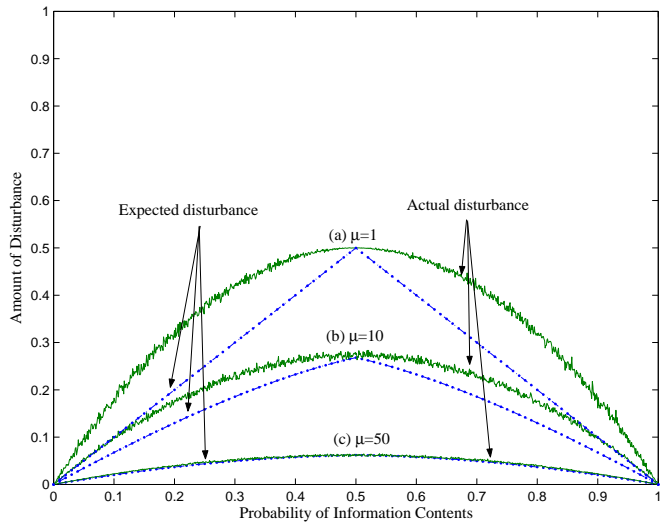}
   \caption{The amount of disturbance introduced to the system using the measurement based quantum random walks 
   with different values of $\mu$, where the solid lines refers to the simulation results and the dotted lines 
   is expected amount of disturbance shown in Eqn.(\ref{psucc}).}
   \label{errorplot}
\end{center}
\end{figure}
\end{center}

For $\varphi>\frac{\pi}{4}$, the same equation (Eqn.(\ref{psucc})) can be used as the probability of success of 
the algorithm but with 
$\Delta J = \sqrt {\frac{2}{\pi }J}$. As an illustrative example, Fig. \ref{readout} shows simulation 
results of the algorithm compared with the probability of success shown in Eqn.(\ref{psucc}) by setting $J=100$ for $\mu=1$, $\mu=10$ and 
$\mu=50$. The simulation results shown in Fig. \ref{readout} is the average of the probability of success 
to read the information of $\left| \psi  \right\rangle$. The simulation results are collected by applying 
the algorithm iteratively for $0 \le \sin ^2 \left( \phi  \right) \le 1$ 
with step 0.001 and each step is repeated 1000 times. Taking the probability of success of $\varphi=0$ as a reference 
probability relevant to the probability of success of projective measurement, 
so iterating the algorithm for 100 items gives a probability of success of 1.0 using $\mu=1$, 0.74224 using $\mu=1$, 
and 0.52233 using $\mu=50$ which is close to a random guess. 

Based on the same example shown in Fig. \ref{readout}, Fig. \ref{errorplot} shows the actual amount of 
disturbance introduced to the system using the proposed algorithm taken as the average disturbance 
from all the trials compared with the expected amount of disturbance $d_e$ calculated as follows

\begin{equation}
 d_e = |\cos^{\rm 2} \left( \phi  \right){\rm  - Pr_j}\left( {\left| {\psi _{\rm 0} } \right\rangle } \right){\rm |}.
\end{equation}

\begin{center}
\begin{figure}[htbp]
\begin{center}
   \includegraphics[width=250pt]{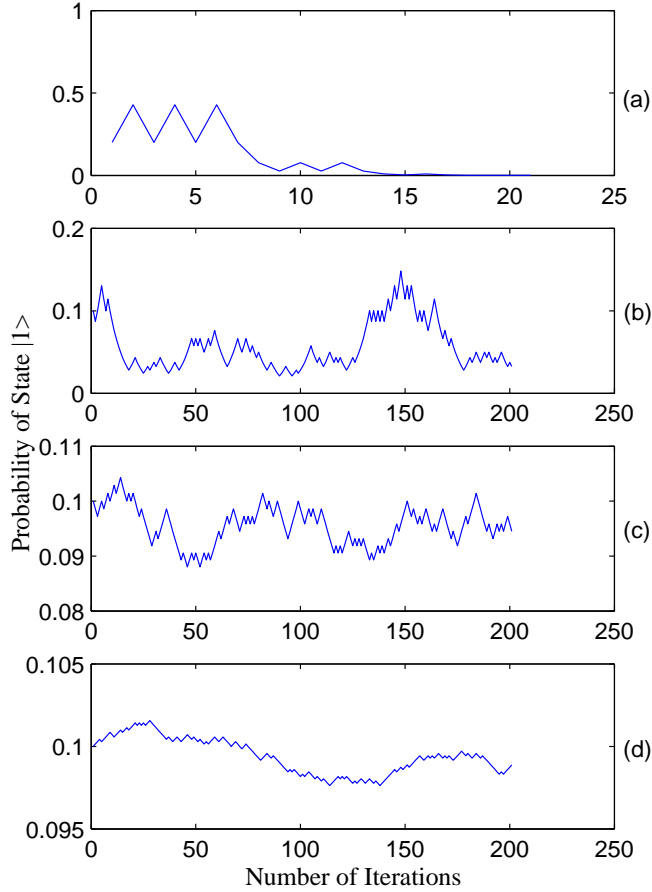}
   \caption{Measurement Based quantum random walks with (a) $\mu=1$, (b)$\mu=10$, (c)$\mu=100$ and $\mu=1000$.}
   \label{qrm1}
\end{center}
\end{figure}
\end{center}

Fig. \ref{qrm1}.(a) shows a MBQRW with $\mu=1$ where the $\left| \psi  \right\rangle$ will collapse to either $\left| 0  \right\rangle$ or 
$\left| 1  \right\rangle$ very fast with probabilities close to $\cos ^2 \left( \phi  \right)$ or 
$\sin ^2 \left( \phi  \right)$ respectively. This gives high accuracy but will disturb the superposition in a way very close 
to the projective measurement.
 
Fig. \ref{qrm1}.(b) shows a MBQRW with $\mu=10$ where $\left| \psi  \right\rangle$ will not collapse to 
$\left| 0  \right\rangle$ or $\left| 1  \right\rangle$ but will make it move up or down 
with probabilities not far from $\cos ^2 \left( \phi  \right)$ or 
$\sin ^2 \left( \phi  \right)$ respectively. This gives acceptable accuracy and will not disturb the superposition very much.

Fig. \ref{qrm1}.(c) and Fig. \ref{qrm1}.(d)  show MBQRWs with large number of dummy qubits $\mu$
where $\left| \psi  \right\rangle$ will not collapse to either $\left| 0  \right\rangle$ or 
$\left| 1  \right\rangle$ and information gain about $\left| \psi  \right\rangle$ will be no better than a random guess.

\section{Conclusion}

In this paper, a quantum algorithm has be proposed 
to read the information content of an unknown qubit without using sharp measurement on that qubit. 
The proposed algorithm used a partial negation operator that creates a weak entanglement between 
the unknown qubit and the an auxiliary qubit. A quantum feedback control scheme is used  
 where sharp measurement is applied iteratively on the auxiliary qubit. 
Counting the outcomes from the sharp measurement on the auxiliary qubit has been used to read 
the information contents on the unknown qubit. It has been shown that the iterative measurements 
on the auxiliary qubit makes the amplitudes of the superposition move in a random walk manner. 
The random walk has a reversal effect when moved in opposite directions, this helps to decrease 
the disturbance that will be introduced to the system during the run of the algorithm. 
The proposed algorithm defined the strength 
of the weak measurement as the distance the random walk has to move from the initial state to the state of 
the sharp measurement which can controlled by using an arbitrary number of dummy qubits $\mu$ in the system.
Adding more dummy qubits to the system made the measurement process slower 
so that the effect of the sharp measurement will be reached after $O(\mu^2)$ measurements on the auxiliary qubit. 
It has been shown that the more we disturb the system, the more information we can get about that system.

\section*{Acknowledgement}
I would like to gratefully thank Prof. Jonathan E. Rowe (University of Birmingham) for his valuable 
comments and suggestions on an earlier version of this work that greatly improved the manuscript.

\end{document}